\documentstyle[12pt]{article}
\newcommand{\la}[1]{\label{#1}}
\newcommand{\beq}{\begin{equation}}
\newcommand{\eeq}{\end{equation}}
\newcommand{\bea}{\begin{eqnarray}}
\newcommand{\ba}{\begin{eqnarray*}}
\newcommand{\eea}{\end{eqnarray}}
\newcommand{\ea}{\end{eqnarray*}}
\newcommand{\mn}{\mu\nu}
\newcommand{\ie}{{\em ie\ }}
\newcommand{\eg}{e.g.\ }

\newcommand{\eq}{eq.~}
\newcommand{\eqs}{eqs.~}
\newcommand{\fig}{fig.~}

\newcommand{\der}[2]{{\frac{{\rm d}#1}{{\rm d}#2}}}
\newcommand{\oder}[2]{{\frac{\partial #1}{\partial #2}}}
\renewcommand{\vec}[1]{{\bf#1}}
\let\oldref=\ref
\renewcommand{\ref}[1]{(\oldref{#1})}
\newcommand{\doo}{\partial}

\newcommand{\Tf}{T_{\rm f}}

\newcommand{\vd}{v_{\rm def}}
\begin{document}
\setcounter{page}{0}
\setlength{\parindent}{0.0cm}
\title{ }
\author{{\bf M. Laine}\thanks{\twlrm email: mlaine@finuhcb}}
\date{ }
\maketitle
\begin{center}
\vspace*{-4cm}
{\Large\bf BUBBLE GROWTH AS A DETONATION} \\
\vspace*{2.5cm}
{\sl  Department of Theoretical Physics}
\\
{\sl  P.O. Box 9, 00014 University of Helsinki, Finland}
\\
\vspace*{0.3cm}
7 September 1993
\end{center}
\vspace*{-8cm}
\hfill Preprint HU-TFT-93-44
\vspace*{8cm}
\vspace*{1.5cm}
\begin{center}
{\large\bf Abstract}\\
\vspace*{5mm}
\parbox{14cm}{
In the case of spherically symmetric chemical burning, only very special
hydrodynamical detonation solutions exist. These are the so called Jouguet
detonations, in which the burning front moves at sound velocity with
respect to the burnt matter. Usually it is believed that the situation
is similar in the case of bubble growth in cosmological phase transitions.
In this paper it is shown that actually a much larger class of detonation
solutions exists in cosmological phase transitions.\\

\vspace*{1mm}
}
\end{center}
\newpage
\setcounter{page}{1}
\setlength{\parindent}{1.2cm}

\section{Introduction}
First order phase transitions in cosmology --- probably at least the
electro\-weak phase transition and the quark-hadron phase transition ---
proceed through the supercooling of the high-temperature phase and the
nucleation of supercritical bubbles of the low-temperature phase.
The nucleated bubbles start to grow. Assuming that this process is
mainly hydrodynamical, the bubbles can grow either as deflagrations
or as detonations. In a large part of parameter space, the bubbles
grow as deflagrations which must be preceded by shock waves to satisfy
the boundary conditions. However, if the supercooling is considerable
the bubbles can grow as detonations~\cite{gyulassy,eikr}.

Both deflagration and detonation solutions can be further subdivided into
three categories which have different qualitative
features~\cite{landau,courant}. The categories are called the strong,
the weak and the Jouguet detonations and deflagrations. These different
types of processes are characterized by the velocity of the combustion
front with respect to the matter behind the front, the Jouguet processes
being those for which matter flows out of the interface at sound velocity.
In any attempt to account for the precise mechanism of a first order
phase transition one should in the very beginning be aware of
the category to which the growing bubbles belong. However, to be able
to do so one must give up the idealization of viewing the phase
transition surface as a discontinuity and to study its microscopic
physics. In the context of chemical burning this problem is well
understood~\cite{landau,courant}. In the case of cosmological phase
transitions, less is known.

In this paper we show what the relevant
categories of detonations are, and we also review the
situation in the case of deflagrations. For chemical burning
taking place as a deflagration, it is well known that strong
deflagrations are impossible. This means that the burning front
cannot move supersonically with respect to the burnt matter. This result
will be restated in section~4 in the context of cosmological phase transitions.
However, nothing more can said about the velocity of the
phase transition surface without microscopic calculations such as those
performed in refs.~\cite{dine}--\cite{xlebnikov} for
the electroweak phase transition  and in ref.~\cite{keijo}
for the quark-hadron phase transition.
For bubbles growing as detonations the situation is different
and quite interesting. Namely, in the case of chemical burning
the bubbles can only grow as Jouguet detonations~\cite{landau,courant}.
This remarkable result (called the Chapman-Jouguet hypothesis)
means in particular that the velocity of the combustion front
is completely determined by energy-momentum conservation and by the
boundary conditions, and no degrees of freedom are left to the
microscopic physics of the burning surface. In the case of
phase transitions, there has been some confusion about the validity
of this result. Steinhardt has argued in ref.~\cite{steinhardt}, using
the same arguments as Landau and Lifshitz used
for chemical burning in ref.~\cite{landau},
that the Chapman-Jouguet hypothesis should be valid. His proof
was originally in 1+3 dimensions but it is very easily ``extended'' to 1+1
dimensions as well. The authors in for instance refs.~\cite{gyulassy,liu}
were also under the impression that only the Jouguet detonations
are possible, though at least in refs.~\cite{eikr,hkllm}
the weak detonations in 1+1 dimensions are accepted as well.
There seems to have prevailed the belief that the dimension of the space
would have something to do with the validity of Steinhardt's results.
In refs.~\cite{kamion,abney} the authors use Steinhardt's results
in the 1+3-dimensional context, and also in ref.~\cite{bonometto} there is
some uncertainty about the possibility of the different kinds of detonations.
It is the purpose of this paper to show that
due to qualitative differences between chemical burning and phase
transitions, the Chapman-Jouguet hypothesis does {\em not} hold
in the case of phase transitions and therefore
weak detonations are possible, in addition to the
Jouguet detonations.

The plan of the paper is as follows. In section 2 the relativistic
hydrodynamics of first order phase transitions is reviewed. In section 3,
it is proved that strong detonations
are not possible as a means of bubble growth in phase transitions.
In section 4, it is shown that strong deflagrations are impossible as well.
In section 5 we address the question of weak detonations and show
that these are a natural means of bubble growth in phase transitions.
The conclusions are in section 6.

\section{The hydrodynamics of bubble growth}
Consider a spherically expanding combustion front. Locally the combustion
front looks planar. If the front velocity is constant, we can Lorentz
transform into the rest frame of the front. In this frame, we denote
the matter ahead of the front (the unburnt or the ``quark'' matter)
with subscript 1 and the matter behind the front (the burnt or the
``hadron'' matter) with subscript 2. It is a very good first idealization
to treat the front as a discontinuity. The strong deviation from
thermodynamical equilibrium attached with the burning of the quark phase
to the hadron phase and all the entropy production is confined to this
discontinuity, and outside it the energy-momentum tensor is that
of an ideal fluid: $T^{\mn}=wu^{\mu}u^{\nu}-pg^{\mn}$. Here $w=e+p$ is
the enthalpy density, $e$ is the energy density, $p$ is the pressure
and $u^{\mu}=(\gamma ,\gamma\vec{v})$ is the four-velocity of the fluid.
The energy-momentum conservation across the discontinuity yields the equations
\beq
\left.
\begin{array}{l}
w_1\gamma ^2_1v_1=w_2\gamma ^2_2v_2
\\
w_1\gamma ^2_1v_1^2+p_1=w_2\gamma ^2_2v_2^2+p_2\,\, .
\end{array}
\right.
\la{cons}
\eeq
The non-negativity of entropy production is expressed as
\beq
s_2\gamma _2v_2\ge s_1\gamma _1v_1
\,\, ,\la{entropy}
\eeq
where $s$ is the entropy density. For generality,
we also assume that there is a conserved quantity (its density
is denoted by $n$), which is in thermodynamical equilibrium
with the rest of the fluid. Then we have the additional equation
\beq
n_1\gamma _1v_1=n_2\gamma _2v_2\equiv j
\,\, .\la{charge}
\eeq
In the quark-hadron phase transition the conserved
quantity is the baryon number. In the electroweak phase transition no such
conserved quantity exists, but one can always add a fictitious conserved
quantity (a ``tracer'') to the fluid as long as it does not show in the
equation of state. To see how this can be done, consider a problem
in which there is no conserved quantity, $\mu =0$.
Then the equation of state is $p=p(T)$ and the complete solution
of the hydrodynamical problem (possibly involving discontinuities) consists
of the functions $T(t,\vec{x})$, $\vec{v}(t,\vec{x})$ and $\phi (t,\vec{x})$.
Here~$\phi (t,\vec{x})$ is the order parameter and it is needed when the
problem contains a phase transition. At the initial time $t=0$, choose
an arbitrary function ${n}(0,\vec{x})$, for instance
${n}(0,\vec{x})=1$ in some units.
Then one can integrate the first order
partial differential equation $\doo _{\mu}({n}u^{\mu})=0$, where $u^{\mu}$
is now a known function of $t$ and $\vec{x}$. The solution
${n}(t,\vec{x})$ is the required conserved quantity. Although $\mu =0$,
the quantity ${n}$ enters our thermodynamical equations, because it is
natural in the present context to formulate the thermodynamical
identities using the entropy per tracer ${\sigma}=s/{n}$.
Then the quantity $x=w/n^2$ will prove to be very useful, see below.
Notice also that the arbitrariness of the initial condition for
${n}(t,\vec{x})$ means just that it is equivalent to say that
the entropy per one tracer is increased, or that the entropy per
ten tracers is increased. Finally, in a thermal fluid with
vanishing chemical potential, the true physical particle density
$\tilde{n}$ of all the massless relativistic particles (and antiparticles)
is proportional to entropy and therefore satisfies the
inequality $\doo _{\mu}(\tilde{n}u^{\mu})\ge 0$ in contrast to the
equation satisfied by the tracer. The inequality sign is obeyed
at least at the discontinuities. In the following, when we speak
of baryons we mean just some conserved quantity, possibly the tracer.

If one does not want to use the tracer in the case $\mu =0$,
it is still possible to repeat the following analysis almost
as such, see ref.~\cite{ruuskanen}.
The quantity $x$ only has to be defined in another way.

To describe the state of the fluid on both sides of the discontinuity,
two intensive thermodynamical variables are needed. It is conventional
to take as these the pressure $p$ and the variable $x=w/n^2$ \cite{landau}.
Then the state of the fluid is represented by a point in the $(x,p)$ plane,
and the above continuity conditions relate the two points. Explicitly,
{}from equations~\ref{cons} and~\ref{charge} we get the equations
\bea
-j^2 & = & (p_2-p_1)/(x_2-x_1)
\la{Taub1}
\\
w_2x_2-w_1x_1 & = & (p_2-p_1)(x_2+x_1)\,\, .
\la{Taub2}
\eea
Given the point $(x_1,p_1)$ and the equations of state of both
phases, the enthalpy $w$ can be expressed in terms of $x$ and $p$
and from equation~\ref{Taub2} the curve $p_2=p_2(x_2)$ can
be solved. This curve is called the detonation adiabat. If
both phases are ascribed the same equation of state, then \eqs\ref{Taub1}
and~\ref{Taub2} do not describe a combustion front but a shock front within one
phase. In this case the curve $p_2=p_2(x_2)$ is called the shock adiabat or
the Taub adiabat.

To illustrate the general structure of the detonation and the shock
adiabats in the $(x_2,p_2)$-plane, we need to know something about the
equations of state of the burnt and the unburnt matter. In view of
the comments after \eq\ref{charge}, we could use for
instance the bag equation of state\footnote{
In the bag equation of state, the pressures and the energy densities
are $p_1=a_1T_1^4-B$, $e_1=3a_1T_1^4+B$, $p_2=a_2T_2^4$ and
$e_2=3a_2T_2^4$. Here $a_1=a_2+B/T_c^4$ and $B$ is the bag constant.}
in this purpose as was done in ref.~\cite{steinhardt}, although this would mean
that the chemical potential vanishes and hence there is no true
conserved physical quantity. However, we can easily work
a bit more generally to understand some of the underlying assumptions.
Consider first the shock adiabat of the unburnt matter through
the point $(x_1,p_1)$. Taking either the equation of state
in ref.~\cite[eq.~(2.3)]{kajantie} with non-zero chemical potential,
or the bag equation of state,
we obtain the relation $w_1=4p_1+4B$. Then from \eq\ref{Taub2}
the equation of the shock adiabat becomes
\beq
[p_s+\frac{1}{3}(p_1+4B)][x_s-\frac{1}{3}x_1]=\frac{8}{9}x_1(p_1+B)
\,\, . \la{hyber}
\eeq
The curve $p_{\rm s}=p_{\rm s}(x_{\rm s})$ is a hyperbola and it is drawn
schematically in figure 1 with dashed line.
Consider then the detonation adiabat. It is essential that the detonation
and the shock adiabats never cross and that the detonation adiabat always
lies above the shock adiabat. To prove the first claim, notice that it follows
{}from equation~\ref{Taub2} that if the two adiabats crossed, then the
final states would have exactly the same enthalpy density, particle
density and pressure. This is not possible because the two phases have a
different equation of state: $w_1(n_2,p_2)\neq w_2(n_2,p_2)$. For instance,
at the critical temperature in the quark-hadron phase transition
we have $s_1(T_c,\mu _c)>s_2(T_c,\mu _c)$
because of the latent heat and $n_1>n_2$ as is
seen in ref.~\cite[fig.~3]{kajantie}, so that $w_1=T_cs_1+\mu _cn_1>w_2$.
Specifically, for the bag equation of state we have $w_1(p_2)-w_2(p_2)=4B$.
To prove the second claim, consider equation~\ref{Taub1}
in the limit that $x_2$ approaches $x_1$ from below.
It is seen that $p_2>p_1$ for all $x_2<x_1$. Therefore, because
the adiabats cannot cross even at $x_2=x_1$, the point $(x_1,p_2)$
must lie above $(x_1,p_1)$ and the claim is proved.
With this knowledge, the general structure of the detonation adiabat
can be drawn. In many cases (\eg in the
bag equation of state in the limit $B\to 0$) it is even so
that tuning some parameters makes the equations of state of the burnt
and the unburnt phases the same so that near this limit the detonation adiabat
resembles greatly the shock adiabat. To prove more
precise statements about the detonation adiabat, one should combine
the non-relativistic analysis in refs.~\cite{landau,courant} with
the relativistic analysis in ref.~\cite{thorne}. The detonation
adiabat is represented schematically in figure~1 with solid~line.

We can now enumerate the different qualitative processes by which
the bubble growth can proceed.
{}From equation~\ref{Taub1} it is known that the line in the $(x,p)$-plane
through points $(x_1,p_1)$ and $(x_2,p_2)$ always has a negative slope.
Therefore, there are two kinds of solutions: either $p_2>p_1$ or
$p_1>p_2$. Solutions of the first kind are detonations; examples
of these are the points~B and~C in figure~1 where the dash-dotted line
intersects the detonation adiabat. Solutions of the second
kind (points~F and~G in figure~1) are deflagrations. There are three kinds
of both detonations and deflagrations. Point C is a {\em strong} detonation,
point B is a {\em weak} detonation and point E is a {\em Jouguet} (or
Chapman-Jouguet) detonation. Similarly, point G is a strong deflagration,
point F a weak deflagration and point H a Jouguet deflagration.
{}From energy-momentum conservation alone none of these processes can be
excluded. Notice, however, that the entropy condition~\ref{entropy} does
in many cases rule out the whole family of detonations~\cite[fig. 17]{eikr}
but it does not categorically rule out any of the different types of
detonations: when the phase transition is preceded by sufficient
supercooling, all the processes are in principle possible.

The different types of processes can be characterized
by the velocities at which matter flows into and
out of the discontinuity. To be able to do so, note first
that the slope of a shock adiabat through any point
in the $(x,p)$-plane is the quantity $-n^2\gamma _s^2v_s^2$, where
$v_s$ is the local sound velocity~\cite{thorne}. Second, consider
the detonation adiabat through the point $(x_2,p_2)$, which is hereafter
called point~2. This curve is {\em not} the same as the shock adiabat
through point~2. It can be seen using \eqs\ref{Taub1} and~\ref{Taub2} and
the relativistic versions in ref.~\cite{thorne} of the results
of ref.~\cite[\S 129]{landau} that the detonation and the shock
adiabats through point~2 cross at exactly two points,
namely the points {\sc B} and {\sc C} in figure~1 when point~2 is either of
these, and that the slopes of the detonation
and the shock adiabat agree only if point~2 is point~{\sc E}.
Therefore, contrary to Steinhardt's claim in ref.~\cite{steinhardt},
the slope of the detonation adiabat does not give the local value of
the quantity $-n^2\gamma _s^2v_s^2$. However, when the slope of the detonation
adiabat is bigger than the slope of the straight line between the points
$(x_1,p_1)$ and $(x_2,p_2)$, then also the slope of the shock adiabat
at point~2 is bigger than the slope of this line, and vice versa.
Because the slope of the straight line between
$(x_1,p_1)$ and $(x_2,p_2)$ is by \eq\ref{Taub1} just $-j^2=
-n_1^2\gamma _1^2v_1^2=-n_2^2\gamma _2^2v_2^2$, we see directly from
figure 1 the following characteristics of the different processes:
\begin{itemize}
\item strong detonation: $v_1>v_{s1}$, $v_2<v_{s2}$

\item Jouguet detonation: $v_1>v_{s1}$, $v_2=v_{s2}$

\item weak detonation: $v_1>v_{s1}$, $v_2>v_{s2}$

\item strong deflagration: $v_1<v_{s1}$, $v_2>v_{s2}$

\item Jouguet deflagration: $v_1<v_{s1}$, $v_2=v_{s2}$

\item weak deflagration: $v_1<v_{s1}$, $v_2<v_{s2}$

\end{itemize}
Here $v_{s1}$ is the local sound speed
of the unburnt matter at the thermodynamical state $(x_1,p_1)$
and $v_{s2}$ is that of the burnt matter at $(x_2,p_2)$.
For instance, a weak detonation is supersonic relative
to the matter both behind and in front of the surface.
To relate the velocities $v_1$ and $v_2$ to the
propagation velocity of the combustion front,
boundary conditions must be considered.
The boundary conditions of an expanding
bubble are that the matter far ahead of the phase transition surface
(where no information of the phase transition has yet arrived) and far
behind the phase transition surface (inside the bubble of the
low temperature phase) is at rest. To satisfy these boundary conditions,
the flow profile of a growing bubble consists of several regions~\cite{hannu}.
In the case of a detonation, the matter ahead of the combustion front
is at rest (see \eg fig.~2). Therefore the velocity of the detonation
front $v_{\rm det}$ is the velocity $v_1$ at which matter flows {\em into}
the combustion front. In the case of a deflagration, the matter
behind the combustion front is at rest (see \eg fig.~3). Therefore
the velocity of the deflagration front $v_{\rm def}$ is the velocity
$v_2$ at which matter flows {\em out of} the combustion front.
Hence, for instance, weak deflagrations move subsonically.

It will be useful to notice that formally,
treating every front as a discontinuity, a detonation front
is equivalent to a shock front which is followed by a deflagration front
moving with the same velocity as the shock front. We prove this for
planar fronts. Consider a process in which the state of the fluid changes
{}from $(x_1,p_1)$ to the intermediate state $(x_0,p_0)$ and then from
$(x_0,p_0)$ to $(x_2,p_2)$. {}From equation~\ref{Taub2} we get
$$
\left.
\begin{array}{lr}
w_0x_0-w_1x_1=(p_0-p_1)(x_0+x_1) & \mbox{  }
\\
w_2x_2-w_0x_0=(p_2-p_0)(x_2+x_0) & \mbox{ .}
\end{array}
\right.
$$
Summing these together we see that the points $(x_1,p_1)$ and $(x_2,p_2)$
satisfy \eq\ref{Taub2} if
$$
\frac{p_0-p_1}{x_0-x_1}=\frac{p_2-p_1}{x_2-x_1}
\,\, .
$$
Therefore, the point $(x_0,p_0)$ is just the point D in figure 1 and
the route {\sc AC} is equivalent to the route {\sc ADC}.
The part {\sc AD} is a shock and the part {\sc DC} a deflagration.
{}From the above it follows that the baryon flux $j$ attains
the same value at the shock front and at the deflagration front.
But then matter flows out of the shock front at the same velocity
as it flows into the deflagration front, which
in the planar case implies that the shock and the deflagration fronts
move with the same velocity. In a similarity solution, which is valid after
the early stages of bubble growth, the fronts
are therefore at the same place. Finally, because matter flows
out of a weak detonation front with supersonic velocity, a {\em weak}
detonation is equivalent to a shock and a {\em strong} deflagration
and a strong detonation is equivalent to a shock and a weak deflagration.

\section{The impossibility of strong detonations}
Of the presented six mechanisms of a phase transition, only some are
actually realized as physical processes. To begin with, an expanding
bubble of the low temperature phase {\em cannot} grow as a strong
detonation. This can be seen even on very general arguments. Basically,
in a strong detonation there are not enough degrees of freedom to
adjust to arbitrary boundary conditions~\cite{courant} and specifically
not to those of an expanding bubble. We will next prove the
impossibility of strong detonations very explicitly, too, since
this proof will provide us with valuable information. For simplicity,
our proof will be in 1+1 dimensions. An analogous proof in 1+3 dimensions
has been given in ref.~\cite{landau} for the non-relativistic case
and with some modifications in ref.~\cite{steinhardt} for the relativistic
case.

The equations governing the evolution of a bubble are the jump conditions
in \eqs \ref{Taub1} and~\ref{Taub2},
the energy-momentum conservation $\doo _{\mu}T^{\mn}=0$
and the charge conservation $\doo _{\mu}(nu^{\mu})=0$.
Denote by $\sigma$ the entropy per baryon.
Projecting the equation $\doo _{\mu}T^{\mn}=0$ in the direction
of and in the direction perpendicular to the flow velocity $u^{\mu}$,
the energy-momentum and charge conservation equations can be written
in the form
$$
\left.
\begin{array}{lr}
\doo _t(n\gamma )+\doo _x(n\gamma v)=0 & \mbox{  }
\\
\doo _t\sigma+v\doo _x\sigma =0 & \mbox{  }
\\
\doo _tv+v\doo _xv=-(\doo _xp+v\doo _tp)/w\gamma ^2 & \mbox{ .}
\end{array}
\right.
$$
For a similarity solution depending only on the variable $\xi =x/t$ these
equations reduce to
\bea
(\xi -v)n' & = & (1-\xi v)n\gamma ^2v' \la{sim1}
\\
(\xi -v)\sigma ' & = & 0 \la{sim2}
\\
(\xi -v)v' & = & (1-\xi v)p'/w\gamma ^2  \la{sim3}
\eea
where the prime denotes differentiation by $\xi$. {}From \eq \ref{sim1} it
follows that the solution cannot be $v=\xi$. Then \eq \ref{sim2} implies
that $\sigma '=0$. But now using simple thermodynamics
we can write $p'$ in terms of $n'$:
$$
\der{p}{\xi}=\left(\oder{p}{e}\right)_{\sigma}
\left(\oder{e}{n}\right)_{\sigma}\der{n}{\xi}=v_s^2(w/n)\der{n}{\xi} \,\, .
$$
Then \eqs \ref{sim1} and~\ref{sim3} give an equation for $v(\xi )$ alone,
which is trivially solved to yield two solutions, namely $v'=0$ and
\beq
v_s(\xi )=\frac{\xi -v(\xi )}{1-\xi v(\xi )}
\,\, . \la{rare}
\eeq
Putting these two types of solutions together, we get a picture
of the bubble growing as a detonation. For $0\le\xi\le v_s$, the
fluid is at rest. At $\xi =v_s$ starts the rarefaction solution,
\eq \ref{rare}. Then there are two possibilities: either
the phase transition discontinuity comes straight after the rarefaction
solution, in which case it follows from \eq \ref{rare} that matter
flows out of the discontinuity at exactly the sound speed, or
there is an area of constant velocity between the rarefaction and
the discontinuity. In the latter case matter flows out of the discontinuity
at a velocity greater than $v_s$ because the right hand side of \eq \ref{rare}
is a growing function of $\xi$ for constant $v$. The former possibility is
a Jouguet detonation, the latter a weak detonation. These are shown
schematically in figure 2, together with the corresponding 1+3 -dimensional
detonation bubbles. Hence, strong detonations are not possible.

\section{The impossibility of strong deflagrations}
Another type of solutions which are not realized in nature is strong
deflagrations. Performing some causal analysis near the interface,
one can conclude that strong deflagrations would not be
mechanically stable and hence they cannot exist~\cite[\S 131]{landau}.
Another way to exclude these is to give up the idealization
of viewing the phase transition surface as a discontinuity and
to study its microscopic structure. The essential
feature which makes strong deflagrations
impossible is that thermodynamical variables --- to the extent that
they can be used in this region --- change monotonically and continuously
between the two phases~\cite{courant}. Then in a strong deflagration the
state of matter would slide from point~{\sc A} in figure~1 to point~{\sc G}.
Assume that inside the burning zone matter can be described as a changing
mixture of the low-temperature phase and the high-temperature phase, so that
the energy-momentum tensor is that of an ideal
fluid\footnote{When there is a microscopic order parameter field,
one has to make the proof somewhat differently.}. Then
equation~\ref{Taub2} is satisfied at every intermediate
point $(\tilde{x}_2,\tilde{p}_2)$ between {\sc A} and {\sc G}.
This means that at point~{\sc F}
the state of the system characterized by $\tilde{x}_2$, $\tilde{p}_2$,
$\tilde{w}_2$ (from eq.~\ref{Taub2}) and $\tilde{n}_2$ (from the
equation $x=w/n^2$) has not only the same variables $x_2$ and $p_2$
as the low-temperature phase at this point but also the same $w_2$ and $n_2$,
since the detonation adiabat is just the solution of eq.~\ref{Taub2}. Because
the equation of state fixes the relation $w_2=w_2(n_2,p_2)$, the system
actually already has to be in the low-temperature phase at point~{\sc F}.
But then the transition from point~{\sc F} to point~{\sc G} would
be completely equivalent to a rarefaction shock, since it was noted
earlier that the shock adiabat through point~{\sc F} crosses the
detonation adiabat also at point~{\sc G}. Since entropy increase does not
allow rarefaction shocks, there cannot be any microscopic mechanism
by which to move from {\sc F} to {\sc G}, \ie strong deflagrations
are not possible. Notice, however, that entropy increase does allow a
similarity rarefaction solution, which is not a discontinuity. Therefore
it seems possible that if a strong deflagration could somehow momentarily be
forced to exist, it would tend to split into a similarity rarefaction solution
and a weak deflagration.

To give an example of a typical deflagration solution,
we cite results from ref.~\cite{ikksl}.
There a simple model is given for first order phase transitions taking
place in relativistic matter. The model includes one phenomenological
parameter $\Gamma$ of dimension $\mbox{GeV}^{-1}$ which is a sort
of dissipation constant and fixes the entropy production at the phase
transition surface. The dissipation constant is roughly proportional
to the collision time $\tau _c$. Small $\Gamma$ means large friction and large
$\Gamma$ small friction at the interface. In this model, it turns out that
in deflagration solutions the temperature and other thermodynamical variables
really behave monotonically at the phase transition surface, see \fig 3.
Accordingly, only weak deflagrations (and as a limiting case
of these, Jouguet deflagrations) have been found.

\section{Weak detonations}
We now turn to weak detonations. It has been argued
in refs.~\cite{landau,courant} for the non-relativistic case of
{\em chemical burning} and in ref.~\cite{steinhardt} for the case
of phase transitions in relativistic matter that weak detonations
are impossible. Then the Jouguet detonation would be left as
the only mechanism for bubble growth as a detonation. This
is called the Chapman-Jouguet hypothesis. To see how this arises,
consider chemical burning. The rate of chemical burning increases
rapidly with increasing temperature, often as $\exp (-U/T)$ where
$U$ is some constant factor. Therefore, heat must be supplied
to the unburnt matter before the burning can begin. In deflagrations,
this happens by thermal conduction from the burning zone where heat
is liberated on account of the exothermic character of the reaction.
This mechanism is, however, so slow that the velocity of the
burning front is subsonic~\cite{landau}. To make a detonation, an entirely
different means of raising the temperature is needed, and this can be achieved
by a strong shock wave which ignites the burning. The shock front compresses
the matter so that the temperature and the pressure rise dramatically, and
then the burning zone follows in which pressure again decreases (the
temperature increases even in the burning zone, which is possible because
the chemical potential is non-vanishing). This means
that the formal equivalence of a detonation to a shock and a deflagration,
which was proved above, would also describe the true microscopic
structure of the detonation front. In figure 1, this means that
matter first jumps from point {\sc A} to point {\sc D} and
then slides from {\sc D}  to {\sc C} as thermodynamical
quantities change continuously in the burning zone. Since a shock
in which pressure decreases is forbidden by the condition of entropy
increase, matter remains in {\sc C} and cannot
continue to {\sc B}. Another way to put this is that a weak detonation
is equivalent to a shock and a strong deflagration, and as strong
deflagrations are impossible, so are weak detonations.
{}From these arguments, the would-be mechanism of a detonation
in chemical burning must be a strong detonation, point {\sc C}.
But because even strong detonations were above proved to be impossible, the
detonation must correspond to point {\sc E} in figure 1. And
this is just the Jouguet detonation, in accordance with
the Chapman-Jouguet hypothesis.

The question is now what the microscopic structure of a detonation
front in the case of a {\em phase transition} actually is. First of all,
no preheating such as that caused by the shock front in the case of
chemical burning is needed to start the phase transition. Namely, in the
case of phase transitions, there is no local ``burning rate'' which
would have to be high enough for the transition to start. The only relevant
rate is the rate of nucleation. This is of the form $\exp (-U/T)$ where
$U$ is not constant but includes singular temperature dependence:
$U\propto 1/(1-T/T_c)^2$~(in the limit of small
supercooling~\cite{eikr,lifshitz}). The lower the temperature is, the faster
is the rate of nucleation. Once the nucleation has happened, the growth
of the bubble is determined by the equation of motion of the order parameter
and by hydrodynamics. It seems highly unlikely that the solution of
these equations would include a sharp narrow peak in pressure,
because due to the vanishingly small chemical potential in the case
of cosmological phase transitions, the peak in pressure would also cause a
sharp narrow peak in temperature. Such a peak in temperature is by itself
rather unnatural, and it would also make the two minima of the effective
potential more degenerate and the potential barrier between the minima higher.
Both factors tend to make the tunneling between the minima more difficult.
Notice also that in the case of deflagrations, a rise in temperature after
the shock front is necessitated by boundary conditions and hydrodynamics,
but the supposed temperature peak at the detonation front serves no
such purpose. For these reasons, it seems much more natural that the order
parameter --- if one exists --- and the thermodynamical variables change
monotonically to their new values in the transition front and the latent heat
released in this process fuels the rapid expansion of the detonation front.
This means that in figure~1 the state of matter changes directly from point
{\sc A} to point {\sc B}. As an example, we quote results from the
above-mentioned model of ref.~\cite{ikksl}. In figure 4, a bubble
growing as a detonation is shown. At the detonation front, the temperature
really changes monotonically and the solution, accordingly, is a weak
detonation. All in all, weak detonations seem to be a natural mechanism for
bubble growth in phase transitions.

To be concrete, let us investigate the quark-hadron phase
transition in the early universe. The physical parameters describing
the first order phase transition are the critical temperature $T_c$,
the latent heat $L$, the surface tension $\sigma$ and the correlation
length $l_c$. We take the values $\sigma =0.1$ and $l_c=6$ in appropriate
powers of $T_c$. For the latent heat we assume a small value $L=0.1T_c^4$,
which results in considerable supercooling so that the nucleation
temperature is $\Tf =0.891T_c$ \cite{keijo}. With this much supercooling,
detonations are possible. With the model of ref.~\cite{ikksl}
and the phenomenological parameter $\Gamma$ describing
the physics of the phase transition surface, figure 5 results.
For small $\Gamma$ the bubbles grow subsonically as weak deflagrations;
for large $\Gamma$ they grow supersonically as weak detonations.
Let us mention in passing that this set of parameters yields a very
interesting scenario for the quark-hadron
phase transition. The average distance between the nucleated bubbles
would be very large, the bubbles would grow as detonations and the
phase transition would leave behind it
large-scale inhomogeneities \cite{ikksl}.

\section{Conclusions}
When entropy increase allows it, a bubble
of the low-temperature phase can grow either as a weak deflagration or a
weak detonation (the Jouguet processes are here considered to be
limiting cases of these). For weak detonations to be possible, the
nucleation must be preceded by considerable supercooling. In weak
deflagrations, the velocity of the phase transition surface is smaller
than or equal to sound velocity, and in weak detonations, it is larger
than sound velocity. In neither case can the energy-momentum conservation
and the boundary conditions alone determine the velocity of the phase
transition surface: its microscopic physics, for instance the entropy
production, must be known before the exact expansion velocity can
be calculated. This means that the detonation solutions are qualitatively
different in the cases of chemical burning and phase transitions.

\section{Acknowledgements}
The author is most grateful to K.~Kajantie for inspiration and support
and to J. Ignatius, H.~Kurki-Suonio and P.~V.~Ruuskanen for useful
discussions.

\newpage
\setlength{\parindent}{0.0cm}
\section*{Figure captions}
{\bf Figure 1:}
\begin{minipage}[t]{13cm}
A schematic picture of the detonation and the shock adiabats. Point {\sc A}
is the initial state $(x_1,p_1)$ of the unburnt fluid. The detonation adiabat
shows the possible final states in the $(x_2,p_2)$-plane.
\end{minipage} \\
\vspace*{0.5cm}

{\bf Figure 2:}
\begin{minipage}[t]{13cm}
On the left is a schematic illustration of the velocity profiles
in a 1+1 --dimensional weak (solid line) and Jouguet (dashed line)
detonation. On the right are the same profiles for the 1+3 -dimensional
case~\cite{hannu}.
\end{minipage} \\
\vspace*{0.5cm}

{\bf Figure 3:}
\begin{minipage}[t]{13cm}
A deflagration solution at time $t=1800$ after the nucleation
in the 1+1 --dimensional model of ref.~\cite{ikksl}. The quantities are
measured in appropriate powers of $T_c$ to make them dimensionless.
The front on the right
is the shock front which has at this time not yet sharpened
to an exact discontinuity, and the front on the left is the deflagration front
where the phase transition takes place. Both fronts are moving to the right.
The velocity of the deflagration front is $\vd =0.46$.
\end{minipage} \\
\vspace*{0.5cm}

{\bf Figure 4:}
\begin{minipage}[t]{13cm}
A detonation solution at time $t=1800$ after the nucleation
in the 1+1 --dimensional model of ref.~\cite{ikksl}. The quantities are
measured in appropriate powers of $T_c$ to make them dimensionless.
The front on the right is the detonation
front where the phase transition takes place,
and the front on the left is the rarefaction wave. Both fronts are moving
to the right. The velocity of the detonation front is $v_{\rm det} =0.72$.
Although the rarefaction wave is very thin in this picture due to the
scale of the $y$-axis, it cannot be approximated as a discontinuity, see
fig.~2.
\end{minipage} \\
\vspace*{0.5cm}

{\bf Figure 5:}
\begin{minipage}[t]{13cm}
The process by which nucleated bubbles grow
in the 1+1 -dimensional model of ref.~\cite{ikksl}
as a function of the parameter $\Gamma$ (in units of $1/T_c$)~\cite{ikksl}.
At approximately $\Gamma =10$ the solution changes from
a weak deflagration to a weak detonation.
\end{minipage}

\end{document}